\documentclass[12pt]{article}% scrbook, scrreprt, scrlettr, scrartcl
\usepackage[cp1251]{inputenc}
\usepackage[english,russian]{babel}
\usepackage{graphicx}
\usepackage{mathtext}
\usepackage{indentfirst}
\usepackage{epsfig,amsmath,amsfonts}

\inputencoding{cp1251}

   \def\d{\delta} 
   
  \def\k{\kappa}  
    
    \def\f{\varphi}
\def\ff{\phi}

   \def\L{\Lambda} 
  \def\P{\Pi}

\def\fr{\frac} \def\dfr{\dfrac} 

\def\beq{\begin{equation}}
\def\eeq{\end{equation}}

\def\bear{\begin{eqnarray}}
\def\eear{\end{eqnarray}}

\def\bea*{\begin{eqnarray*}}
\def\eea*{\end{eqnarray*}}

\def\mc{\mathcal}

\def\Tr{\mbox{Tr}}
\def\nn{\nonumber}

\newcommand{\Int}[2]{\int_{#1_{(#2)}}}

\begin{document}

%\pagestyle{headings}%empty, plain, headings
%\pagenumbering{arabic}%arabic,roman, alph
%\renewcommand{\contentsname}{}
\renewcommand{\refname}{\begin{center}References\end{center}}

\title{\bf An exact statement for Wilsonian and Holographic renormalization group}\author{{\bf E.T. Akhmedov}\footnote{akhmedov@itep.ru}{ } and
{\bf E.T. Musaev}\footnote{musaev@itep.ru}\vspace{5mm}\\
{B.Cheremushkinskaya, 25, ITEP, 117218, Moscow, Russia} \\ {and}
\\
{Moscow Institute of Physics and Technology, Dolgoprudny, Russia}}\date{}
\maketitle

\begin{center}{\bf Abstract}\end{center}
We show that Polchinski equations in the D--dimensional
matrix scalar field theory can be reduced at large $N$ to the Hamiltonian equations in a (D+1)-dimensional
theory. In the subsector of the $\Tr \phi^l$ (for all $l$) operators we find the exact
form of the corresponding Hamiltonian. The relation to the Holographic renormalization
group is discussed.

\vspace{10mm}

%\tableofcontents
{\bf 1. Introduction.}
Wilsonian renormalization group \cite{Wilson:1973jj} appears to be a useful
tool in the study of various phenomena in QFT and statistical physics. A convenient
form of the Wilsonian renormalization group is given by the Polchinski equations \cite{Polch}.
Usually these equations are formulated in the scalar field theory.

Some time ago, after the discovery of the AdS/CFT--correspondence \cite{Maldacena:1997re},
it was recognized \cite{Akhmedov:1998vf} that renormalization group equations on
the CFT side are represented by classical equations of motion on the AdS side.
The idea was further developed in \cite{Balasubramanian:1999jd}--\cite{Akhm}.
And it became clear that the holography is the general phenomenon for the proper formulation
of the Wilsonian renormalization group at large $N$ \cite{Mironov:2000ij}, \cite{Akhm}.

In this picture an exact, simple and easy testable statement was missing. The goal of this note
is to provide such a statement. We show that Polchinski equations in the D--dimensional
matrix scalar field theory can be reduced at large $N$ to the Hamiltonian equations in a (D+1)-dimensional
theory. In the subsector of the $\Tr \phi^l$ (for all $l$) operators we find the exact
form of the corresponding Hamiltonian. In concluding section
the relation to the Holographic renormalization group is discussed.

{\bf 2. From renormalization group to the Hamiltonian flow.}
In this section we consider Euclidian D-dimensional matrix
scalar field theory:

\begin{equation}
 \label{1.1}
\mc{S}[\ff]=-\fr{N}{2}\int_p\Tr\left[\ff(p)\,(p^2+m^2)^{\phantom{\frac12}}K_{\L}^{-1}(p^2)\,\ff(-p)\right] + N \mc{S}_I[\ff],
\end{equation}
whose action is written here in the Fourier transformed form.
Here $\phi = ||\phi^{ij}||, \, \, i,j=1,\dots, N$ is the Hermitian matrix;
the function $K_{\L}$ is:

\begin{equation}
K_{\L}(x)=
\begin{cases}
 1, & \mbox{when } x<\L^2;\\
 0, & \mbox{when } x>\L^2,
\end{cases}
\end{equation}
and is quickly changing near the point $x=\L^2$, i.e. $\L$ is an UV cutoff in our theory;
$\mc{S}_I$ is the interaction part of the action, which includes sources as well. In this note
we take:

\begin{equation}
 \label{1.2}
\mc{S}_I[\ff]=\sum_{l=0}^{\infty}\int_{k_1\ldots{}k_l}\,\Tr \left[\ff(k_1)^{\phantom{\frac12}}\ldots\,\ff(k_l)\right]\,J_l(-k_1-\ldots-k_l),
\end{equation}
which is just the Fourier transform of $\sum_{l=0}^\infty \int J_l(x)\, \Tr \phi^l(x)$ representing
the subspace of the complete OPE basis of the theory. The crucial observation for our further
considerations is that in the Fourier transformed form $J_l$ depends only on the sum
of $k$'s --- arguments of $\phi$'s under the traces. If we were considering
operators containing derivatives (e.g. $\Tr \left[\left(\partial_\mu \phi \right)\, \phi^l \,
\left(\partial_\nu \phi\right) \,\phi^{n-l}\right]$),
then the corresponding Fourier transformed sources (e.g. $J^n_{\mu\nu}(-\sum k) \,k^\mu_1\, k^\nu_{l+2}$)
in general would depend on all $k$'s separately.

The Polchinski equation for the theory in question is given in the Appendix (see eq. (\ref{a8})).
Taking the quantum average of it, we arrive at:

\begin{equation}
 \label{1.3}
\left\langle\L\fr{d\mc{S}_I[\ff]}{d\L}\right\rangle = -\fr{1}{2}\int_p\fr{1}{p^2+m^2} \, \L\, \fr{dK_{\L}(p^2)}{d\L} \, \left\langle{}\left[N^{-1}\fr{\d^2\mc{S}_I[\ff]}{\d\ff^{ij}(-p)\d\ff^{ji}(p)} + \fr{\d\mc{S}_I[\ff]}{\d\ff^{ij}(p)}\fr{\d\mc{S}_I[\ff]}{\d\ff^{ji}(-p)}\right]\right\rangle.
\end{equation}
The average is taken over the high--momentum modes only. It means that one should represent $\ff(p)$ as the sum of the high--momentum and low--momentum modes $\ff(p)=\ff_0(p)+\f(p)$
and integrate out the field $\f(p)$ \cite{Wilson:1973jj}. Here $\phi_0(p)$ is the solution of the equations of motion following from the action (\ref{1.1}). As we will see below, taking the expectation value in (\ref{1.3}) is necessary to close the system of equations for the sources \cite{Akhm}.

It is easy to verify the following relations:

\begin{eqnarray}
 \label{1.4}\nn
&&\Tr\left[\fr{\d \mc{S}_I}{\d\ff(p)}\fr{\d \mc{S}_I}{\d\ff(-p)}\right] = \sum_{n,l=0}^{\infty} \int_{p_1\ldots{}p_nk_1\ldots{}k_l} (n\cdot{}l) \, \Tr\left[\ff(p_1)\,\ldots\,\ff(p_n)^{\phantom{\frac12}}\ff(k_1)\,\ldots\,\ff(k_l)\right]\times\\\nn
&&\times{} J_\k(-p-p_1-\ldots-p_n)\,J_\k(p-k_1-\ldots-k_l);\\
&&\\\nn
&&\Tr\left[\fr{\d^2\mc{S}_I}{\d\ff(p)\d\ff(-p)}\right] = \sum_{n=0}^{\infty} \int_{p_1\ldots{}p_n} n \, \sum_{m=0}^{n} \, \Tr\left[\ff(p_1)^{\phantom{\frac12}} \ldots \, \ff(p_m)\right]\, \Tr\left[\ff(p_{m+1})^{\phantom{\frac12}} \ldots \,\ff(p_n)\right] \times\\\nn
&&\times{}J_\k(-p_1-\ldots-p_n).
\end{eqnarray}
Now one should calculate the quantum average of these traces. As usual this is a complicated problem, but there is a way to simplify the final expressions.

First, let us introduce the following notations:

\begin{eqnarray}
 \label{1.5}\nn
	&& \int_{p_{(n)}}:=\int_{p_1\ldots{}p_n}; \\
	&& T_n(p_1,\dots,p_n):=\Tr\left[{\ff_0}(p_1)^{\phantom{\frac12}}
           \ldots\,{\ff_0(p_n)}\right]; \\ \nn
	&& J_l(-k_{(l)}):=J_l(-k_1-\ldots-k_l).
\end{eqnarray}
In these notations the action (\ref{1.2}) takes a short form $\mc{S}_I[\ff_0]=\sum_{l=0}^{\infty}\int_{k_{(l)}} T_l(k_1,\dots, k_l)\, J_l(-k_{(l)})$.
The quantum average of the trace $\Tr\left[({\ff_0}_1+\f_1)\ldots({\ff_0}_n+\f_n)\right]$ over the high--momentum modes can be reduced to the action of some operator on $T_l(k_1,\dots, k_l)$:

\begin{equation*}
\begin{split}
\left\langle\int_{p_{(n)}}\Tr\left[({\ff_0(p_1)}+\f(p_1))^{\phantom{\frac12}} \ldots \, ({\ff_0}(p_n)+\f(p_n))\right]\right\rangle \equiv \nonumber \\
\equiv \int_{p_{(n)}}\int\mc{D}\f{}\,e^{S_0}\,\Tr\left[({\ff_0}(p_1)+\f(p_1))^{\phantom{\frac12}}
\ldots \, ({\ff_0}(p_n)+\f(p_n))\right]=\\
=\int_{p_{(n)}}\int\mc{D}\,\f{}\,e^{S_0}\,\exp\left[\int_p\f_p\fr{\d}{\d\ff_0(p)}\right]\,
\Tr\left[\ff_0(p_1)^{\phantom{\frac12}} \ldots\,\ff_0(p_n)\right] = \nonumber \\ = \hat{W}\left\{\int_{p_{(n)}}\,\Tr\left[{\ff_0}_1^{\phantom{\frac12}} \ldots\,{\ff_0}_n\right]\right\}
=\hat{W}\left[\int_{p_{(n)}}T_n(p_1,\dots,p_n)\right],
\end{split}
\end{equation*}
where $S_0 = -\fr{N}{2}\int_p\Tr[\ff(p)(p^2+m^2)K_{\L}^{-1}(p^2)\,\ff(-p)]$ and

\begin{equation}
 \label{1.7}
\hat{W}=\exp\left(\fr{1}{2N}\int_p\Tr\left[\fr{\d}{\d{\ff_0}_p}\fr{\d}{\d{\ff_0}_{-p}}\right]G_\L(p)\right)
\end{equation}
with $G_\L(p) = K_\L(p^2)/(p^2+m^2)$ being the free propagator\footnote{Using the equation (see e.g. \cite{Sham})

\begin{multline}
\Tr\left[\fr{\d}{\d{\ff_0}_p}\fr{\d}{\d{\ff_0}_{-p}}\right] = \Int{k}{n}\sum_{l,m=1}^{\infty}(l\cdot{}m)T_{l+m-2}(\{k_{l-1}\},\{q_{m-1}\}) \fr{\d^2}{\d{}T_l(\{k_{l-1}\},p)\d{}T_m(\{q_{m-1}\},-p)} + \nonumber \\
+\Int{k}{l-2}\sum_{l=2}^{\infty}{}l\sum_{m=0}^{l-2}T_m(\{k\})T_{l-m-2}(\{k\}) \fr{\d}{\d{}T_l(\{k_{l-2}\},p,-p)} \nonumber \\
\end{multline}
the operator $\hat{W}$ can be written in terms of derivatives
with respect to the natural variables $T_n(k_1,\dots,k_n)$
Which makes it obvious that such actions as (\ref{1.2})
(with single--trace operators only) give raise to multi--trace
operators in the Polchinski equation.}.

We work in the large $N$ limit, where the
following factorization property is in effect:

\begin{equation}
 \label{1.9}
\left\langle \prod_n \Tr O_n\right\rangle=\prod_n\langle \Tr O_n\rangle.
\end{equation}
Using this property and the notation $\tilde{T} = \hat{W}\, T$, we can write

\begin{eqnarray}
 \label{1.10}
\hat{W}[T_l(k_1,\dots,k_l)\,T_n(p_1,\dots,p_n)] = \hat{W}[T_l(k_1,\dots,k_l)]\,
\hat{W}[T_n(p_1,\dots,p_n)] = \nonumber \\ = \tilde{T}_l(k_1,\dots,k_l)\,\tilde{T}_n(p_1,\dots,p_n)
\end{eqnarray}
and the Polchinski equation for the theory (\ref{1.1}) acquires the form:

\begin{eqnarray}
 \label{1.11}
\sum_{l=1}^{\infty}\int_{k_{(l)}}\tilde{T}_{l}(\{k_l\})\dot{J}_l(-k_{(l)}) = \nonumber \\
- \fr{1}{2}\int_p\fr{1}{p^2+m^2}\dot{K}_\L(p^2) \, \left[ N^{-1}\, \sum_{a=1}^{\infty} \sum_{s=0}^{a-1} \int_{k_{(a-1)}}(a+1) \tilde{T}_{a-s-1}(\{k_{s+1}\}) \, \tilde{T}_s(\{k_s\}) \, J_{a+1}(-k_{(a+1)})  \right.\nonumber \\ \left. + \sum_{l,j=1}^{\infty} \int_{q_{(j-1)}k_{(l-1)}} (l\cdot{}j) \, \tilde{T}_{l+j-2}(\{k_{l-1}\},\{q_{j-1}\})J_l(-k_{(l-1)}-p)J_j(-q_{(j-1)}+p)\right],
\end{eqnarray}
where the overdot means the differentiation with respect to $d/d \log \L$.
Note that $\tilde{T}$ depends on $\L$, because the $\hat{W}$ operator does depend on the cut--off.

It will become clear in a moment that the structure of the theory in question
suggests to introduce the momentum conjugate to $J_l(k)$ as follows:

\begin{equation}
\label{1.12}
\P_l(k)=N^{-1}\int_{k_{(l)}}\d^{(D)}\left[k-k_{(l)}\right]\tilde{T}_l(k_1,\dots,k_l).
\end{equation}
This definition reflects the fact that our sources depend only on the
sum of the arguments of $\tilde{T}$'s. And the factor of $N$ was included to make the sources $J_l$ and the canonical momenta $\P_l$ to be of the same order as $N\to\infty$. In these variables eq. (\ref{1.11})
reduces to:

\begin{eqnarray}
\int_q\sum_{l=0}^{\infty}\P_l(q)\dot{J}_l(-q) = -\fr{1}{2}\int_p \frac{\dot{K}_\L(p^2)}{p^2 + m^2} \left[\int_{q_1q_2}\sum_{l,s=0}^{\infty}(l+s+2)\P_l(q_1)\P_s(q_2)J_{l+s+2}(-q_1-q_2)+\right. \nonumber \\
\left.+\int_{q_1q_2}\sum_{f,h=1}^{\infty}(f\cdot{}h)\P_{f+h-2}(q_1+q_2)J_f(-q_1)J_h(-q_2)\right].
 \label{1.13}
\end{eqnarray}
Then the corresponding equations for the sources and for the momenta
can be represented in the form of the Hamiltonian equations:

\begin{equation}
 \label{1.14}
\begin{split}
\frac{d\,J_l(-q)}{dT}&=\fr{\d{}H}{\d\P_l(q)}\\
\frac{d\,\P_l(q)}{dT}&=-\fr{\d{}H}{\d{}J_l(-q)},
\end{split}
\end{equation}
where the ``time'' $dT = d\log\L\, \int_p \frac{\dot{K}_\L(p^2)}{p^2 + m^2}$
is related to the cutoff scale.
The first equation in (\ref{1.14}) follows from the
Polchinski equation\footnote{Note that to make the transformation
from (\ref{1.13}) to the first equation in (\ref{1.14}) legal one has to extend the collection of couplings $J_l\,{\rm Tr}\phi^l$ to the full OPE basis in the theory, then perform the same transformations as we
did to arrive at (\ref{1.14}) \cite{Akhm}. At the end one has to put all the additional sources to zero to obtain (\ref{1.14}) in its present form.} (\ref{1.13}). Recall that this equation imposes the condition that the functional integral of the theory in question is independent
of the cutoff. The second equation in (\ref{1.14}) similarly follows from the derivation of the VEV $\langle \Tr \phi^l(x)\rangle$ with respect to $\L$. Or it may be obtained via the variation of the Polchinski equation with respect to $J_k(-q)$.
The easiest way to see the latter fact is to recall that the effective actions expressed through the
sources and through the VEV's are related to each other via the Legendre (functional Fourier) transformation \cite{Akhm}.

The Hamiltonian can be calculated exactly and has a remarkably simple form:

\begin{equation}
\label{1.15}
\boxed{\begin{aligned}
H = - \fr{1}{2} \, \int_{q_1q_2} \sum_{l,s=0}^{\infty}
\left[(l+s+2)\P_l(q_1)\P_s(q_2)J_{l+s+2}(-q_1-q_2)^{\phantom{\frac12}} + \right.\\
\left.\phantom{\P^{\frac12}}+(l+1)(s+1)\P_{l+s}(q_1+q_2)J_{l+1}(-q_1)J_{s+1}(-q_2)\right].
\end{aligned}}
\end{equation}
We emphasize that the momentum $\P_l(q)$ contains all powers of the traces since it is the result of
the action of the $\hat{W}$ operator on $T_l(k)$.

The trivial observation here is that the summation over $s$ and $l$ in the Hamiltonian
in question can be converted into the integration over the new artificial coordinate
$\sigma \in [0,\pi]$. Such a conversion obviously could have been done in the original action
(\ref{1.1})--(\ref{1.2}).

{\bf 3. Discussion and acknowledgments.}
Thus, we have managed to rewrite the Polchinski equations for the matrix scalar field theory
at large $N$ as the Hamiltonian equations. The configuration space of the obtained $D+1$--dimensional
theory consists of the single trace operators of the original $D$--dimensional theory.
Our result does not depend on whether the theory
in question is renormalizable or not, or even whether it is UV divergent or not.

Why such a relation is important? First of all, it clearly shows
that if one keeps all sources for a subsector of the full OPE basis, then the renormalization
group becomes holographic. Indeed, knowing the values of $J$'s and $\P$'s at some energy scale,
one can find them, through the Hamiltonian equations, at any other scale.

Furthermore, despite the fact that we average over the Gaussian
quadratic part of the action in the transformation from (\ref{1.3}) to (\ref{1.14})-(\ref{1.15}) we still have the complete knowledge of the renormalization group flow in the subsector of the theory in question.
In particular, if one would like to know e.g. where does the UV theory

\begin{eqnarray}
\mc{S}[\ff] = -\fr{N}{2}\int_p \Tr\left[\ff(p)\,(p^2+m^2)^{\phantom{\frac12}}K_{\L}^{-1}(p^2)\,\ff(-p)\right] - \nonumber \\ -
g \, N\, \int_{k_1,\dots, k_4}
\delta\left(\sum_{i=1}^4 k_i\right) \, \Tr \left[\phi(k_1) \,\phi(k_2)\, \phi(k_3)\, \phi(k_4)\right],
\end{eqnarray}
($g=const$) flow under the renormalization group, he just has to solve the Hamiltonian equations
(\ref{1.14}) with the initial conditions $J_4=g$ and $J_n=0$ for all $n\neq 4$ as $\Lambda\to\infty$.

Fortunately enough the subsector of the OPE basis,
which we are considering in this note, factors (at large $N$) under the renormalization group flow
from the rest of the OPE basis. So far we did not find the closed form Hamiltonian if the
other parts of the OPE basis are included into the renormalization group dynamics. This remains to be a challenge for the future work. As well it would have been interesting to reconcile our observations
with the information theory interpretation of the renormalization group flow \cite{Apenko:2009kq}.

We would like to thank A.Mironov and A.Morozov for the valuable discussions.
AET would like to specially thank A.Gerasimov for shearing his ideas, for the
collaboration and initiation of this project. AET would like to thank
MPI--AEI, Golm,  where this work was completed, for the hospitality.

\appendix

\vspace{5mm}

{\bf Appendix.} Here we derive Polchinski equation \cite{Polch} in the case of the D--dimensional
matrix scalar field theory (\ref{1.1}). In Wilsonian renormalization group
one integrates out the high--momentum modes in $\ff$ so that the energy scale is reduced from the cutoff
to a much lower scale $\L_R$ --- the scale, where we are probing our physics \cite{Wilson:1973jj}. We assume that $m^2\ll\L_R^2$.

Consider generating functional for the theory (\ref{1.1}):

\begin{equation}
 \label{a6}
\mc{Z} = \int\mc{D}\ff \, e^{\mc{S}[\ff,\L, \{J\}]}.
\end{equation}
Obviously one has to impose the condition that physics shouldn't depend on the cutoff:

\begin{equation}
\L\fr{d\mc{Z}}{d\L}=0.
\end{equation}
The result of the differentiation is:

\begin{equation}
 \label{a7}
\L\fr{d\mc{Z}}{d\L}=\int\mc{D}\ff e^{S[\ff,\L, \{J\}]} \Tr \left[\int_p\ff(-p)(p^2+m^2)\ff(p)\L\fr{dK_\L^{-1}(p^2)}{d\L} + \L\fr{d\mc{S}_I}{d\L}\right].
\end{equation}
It is easy to verify that the expression under the integral on the RHS of (\ref{a7}) 
becomes full functional derivative if

\begin{equation}
 \label{a8}
\L\fr{d\mc{S}_I}{d\L}=-\fr{1}{2}\int_p\fr{1}{p^2+m^2}\L\fr{dK_{\L}(p^2)}{d\L} \left(\fr{\d^2\mc{S}_I}{\d\ff^{ij}(-p)\d\ff^{ji}(p)} + \fr{\d\mc{S}_I}{\d\ff^{ij}(p)}\fr{\d\mc{S}_I}{\d\ff^{ji}(-p)}\right).
\end{equation}
Indeed substituting (\ref{a8}) into (\ref{a7}) we obtain

\begin{equation}
 \label{a9}
\L\fr{d\mc{Z}}{d\L}=\int_p\L\fr{dK_\L(p^2)}{d\L}\int\mc{D}\ff\,\Tr
\fr{\d}{\d\ff}\,\left[\left(\ff(p)K_\L^{-1}(p^2) + \fr{1}{2}(p^2+m^2)^{-1}\fr{\d}{\d\ff}\right)e^{S[\ff,\L, \{J\}]}\right].
\end{equation}
It is straightforward to see that (\ref{a9}) and (\ref{a7}) are equivalent because:

\begin{eqnarray}
\label{a10}\nonumber
&\dfr{\d{}e^{S[\ff,\L]}}{\d\ff(-p)} = &\left(-\ff(p)(p^2+m^2)K_\L^{-1}(p^2) + \fr{\d\mc{S}_I}{\d\ff}\right)e^{S[\ff,\L, \{J\}]};\\
&\dfr{\d^2e^{S[\ff,\L]}}{\d\ff(-p)\d\ff(p)} = &\left[-(p^2+m^2)K_\L^{-1}(p^2) + \fr{\mc{S}_I}{\d\ff(-p)\d\ff(p)}+\right.\\\nonumber
&&+\left. \left(-\ff(p^2+m^2)K_\L^{-1}(p^2) + \dfr{\d\mc{S}_I}{\d\ff}\right)^2\right]e^{S[\ff,\L, \{J\}]}.
\end{eqnarray}
Eq. (\ref{a8}) is referred to as the Polchinski equation.

\end{document}